\documentclass[runningheads,a4paper]{llncs}
\usepackage{amsmath}
\usepackage{amssymb}
\usepackage{graphicx}
\usepackage{booktabs}
\usepackage{listings}
\usepackage{xcolor}
\usepackage{hyperref}
\usepackage{caption}
\usepackage{subcaption}

\usepackage{url}
\usepackage{color}
\usepackage{cite}
\usepackage[nomargin,inline,marginclue,draft]{fixme}
\usepackage{cleveref}
\fxsetup{status=draft}
\fxsetup{theme=color, mode=multiuser} \FXRegisterAuthor{me}{ame}{\color{red}Me}
\newcommand{\orcid}[1]{\href{https://orcid.org/#1}{\includegraphics[scale=0.02]{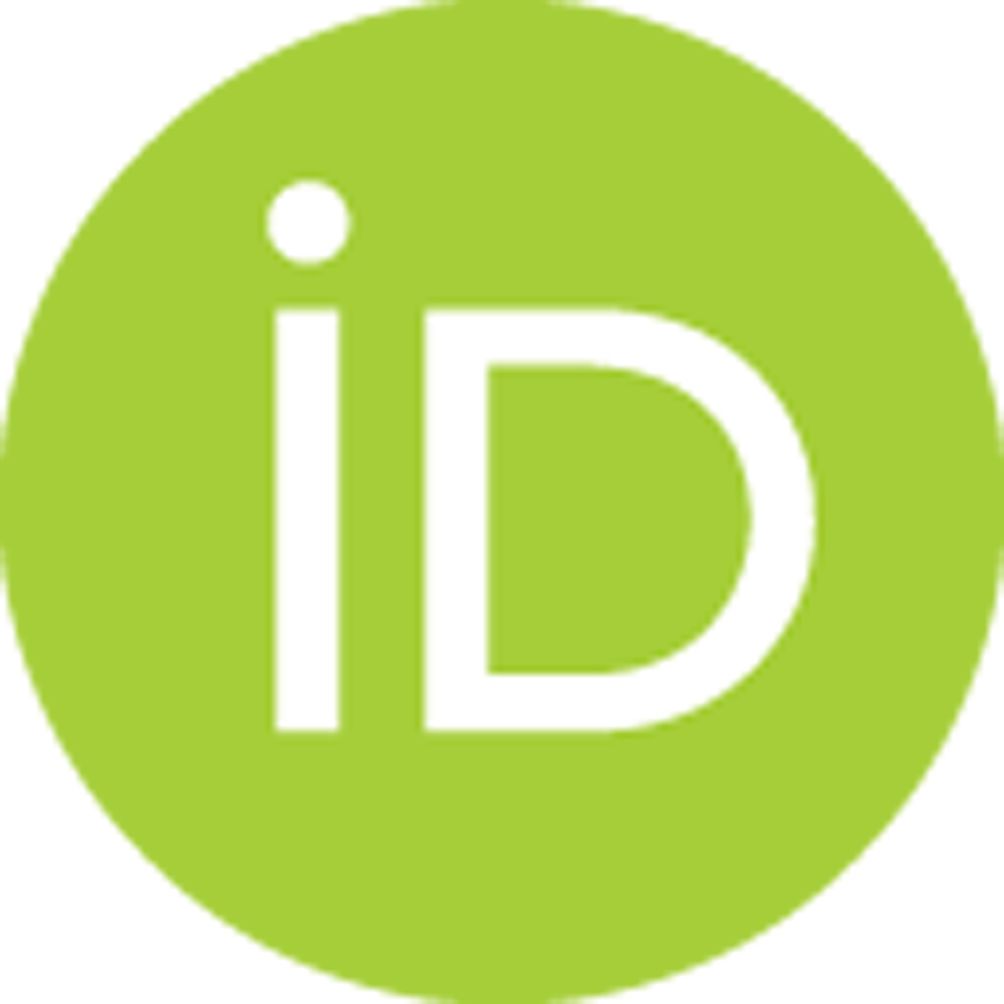}}} % this does not print the whole orcid number

\hypersetup{
    colorlinks=true
}
\title{The Federated Tumor Segmentation (FeTS) Challenge}
\titlerunning{FeTS Challenge}
\author{
Sarthak Pati\inst{1,2,3,4,\S,\dag,\orcid{0000-0003-2243-8487}}
\and
Ujjwal Baid\inst{1,2,3,\S,\dag,\ddag,\orcid{0000-0001-5246-2088}}
\and
Maximilian Zenk\inst{5,\S,\dag}
\and
Brandon Edwards\inst{6,\dag,\orcid{0000-0002-0957-9149}}
\and
Micah Sheller\inst{6,\dag,\orcid{0000-0002-6571-0850}}
\and
G. Anthony Reina\inst{6,\dag,\orcid{0000-0001-9623-9259}}
\and
Patrick Foley\inst{6,\dag,\orcid{0000-0001-9401-3088}}
\and
Alexey Gruzdev\inst{6,\dag,\orcid{0000-0003-3924-4531}}
\and
Jason Martin\inst{6,\dag}
\and
Shadi Albarqouni\inst{4,\dag,\orcid{0000-0003-2157-2211}}
\and
Yong Chen\inst{8,\dag,\orcid{0000-0003-0835-0788}}
\and
Russell Taki Shinohara\inst{1,8,\dag,\orcid{0000-0001-8627-8203}}
\and
Annika Reinke\inst{7,\dag,\orcid{0000-0003-4363-1876}}
\and
David Zimmerer\inst{5,\dag}
\and
John B. Freymann\inst{11,12,\ddag}
\and
Justin S. Kirby\inst{11,12,\ddag}
\and
Christos Davatzikos\inst{1,2.\ddag}
\and
Rivka R. Colen\inst{13,22,\ddag}
\and
Aikaterini Kotrotsou\inst{13,\ddag}
\and
Daniel Marcus\inst{14,15\ddag}
\and
Mikhail Milchenko\inst{14,15,\ddag}
\and
Arash Nazeri\inst{15,\ddag}
\and
Hassan Fathallah-Shaykh\inst{16,\ddag}
\and
Roland Wiest\inst{17,18,\ddag}
\and
Andras Jakab\inst{19,\ddag}
\and
Marc-Andre Weber\inst{20,\ddag}
\and
Abhishek Mahajan\inst{21,\ddag}
\and
Lena Maier-Hein\inst{7,\dag,\orcid{0000-0003-4910-9368}}
\and
Jens Kleesiek\inst{10,\dag}
\and
Bjoern Menze\inst{4,9,\dag,\orcid{0000-0003-4136-5690}}
\and
Klaus Maier-Hein\inst{5,\dag,\orcid{0000-0002-6626-2463}}
\and
Spyridon Bakas\inst{1,2,3,\dag,\ddag,*,\orcid{0000-0001-8734-6482}}
}
\authorrunning{Pati et al.}

\institute{\scriptsize{Center for Biomedical Image Computing and Analytics (CBICA), University of Pennsylvania, Philadelphia, PA, USA
\and
Department of Radiology, Perelman School of Medicine, University of Pennsylvania, Philadelphia, PA, USA
\and
Department of Pathology and Laboratory Medicine, Perelman School of Medicine, University of Pennsylvania, Philadelphia, PA, USA
\and
Department of Informatics, Technical University of Munich, Munich, Germany
\and
Division of Medical Image Computing, German Cancer Research Center (DKFZ), Heidelberg, Germany
\and
Intel Corporation, Santa Clara, CA , USA
\and
Division of Computer Assisted Medical Interventions, DKFZ, Heidelberg, Germany
\and
Department of Biostatistics, Epidemiology and Informatics, Perelman School of Medicine, University of Pennsylvania, Philadelphia, PA, USA
\and
Department of Quantitative Biomedicine, University of Zurich, Zurich, Switzerland
\and
Translational Image-guided Oncology, Institute for AI in Medicine (IKIM), University Hospital Essen, Germany
\and
Leidos Biomedical Research, Inc., Frederick National Laboratory for Cancer Research, Frederick, MD 21701, USA
\and
Cancer Imaging Program, National Cancer Institute, National Institutes of Health, Bethesda, MD 20814, USA
\and
Department of Diagnostic Radiology, University of Texas MD Anderson Cancer Center, Houston, TX, USA
\and
Neuroimaging Informatics and Analysis Center, Washington University, St. Louis, MO, USA
\and
Department of Radiology, Washington University, St. Louis, MO, USA
\and
Department of Neurology, The University of Alabama at Birmingham, Birmingham, AL, USA
\and
Institute for Surgical Technology and Biomechanics, University of Bern, Bern, Switzerland
\and
Support Centre for Advanced Neuroimaging Inselspital, Institute for Diagnostic and Interventional Neuroradiology,Bern University Hospital, Bern, Switzerland
\and
Center for MR-Research, University Children’s Hospital Zurich, Zurich, Switzerland
\and
Institute of Diagnostic and Interventional Radiology, Pediatric Radiology and Neuroradiology, University Medical Center Rostock, Ernst-Heydemann-Str. 6, 18057 Rostock, Germany
\and
Tata Memorial Centre, Homi Bhabha National Institute, Mumbai, India
\and
Hillman Cancer Center, University of Pittsburgh Medical Center, Pittsburgh, PA, 15232, USA}
\\
\textsuperscript{\S} Equally contributing authors.\\ 
\textsuperscript{\dag} People involved in the organization of the challenge.\\
\textsuperscript{\ddag} People contributing data from their institutions.\\
\textsuperscript{*} Corresponding author: \email{\{sbakas@upenn.edu\}}}

\begin{document}
\mainmatter
\maketitle
\newpage
\setcounter{footnote}{0} 
\begin{abstract}
    This manuscript describes the first challenge on Federated Learning, namely the Federated Tumor Segmentation (FeTS) challenge 2021. International challenges have become the standard for validation of biomedical image analysis methods. However, the actual performance of participating (even the winning) algorithms on ``real-world'' clinical data often remains unclear, as the data included in challenges are usually acquired in very controlled settings at few institutions. The seemingly obvious solution of just collecting increasingly more data from more institutions in such challenges does not scale well due to privacy and ownership hurdles. Towards alleviating these concerns, we are proposing the FeTS challenge 2021 to cater towards both the development and the evaluation of models for the segmentation of intrinsically heterogeneous (in appearance, shape, and histology) brain tumors, namely gliomas. Specifically, the FeTS 2021 challenge uses clinically acquired, multi-institutional magnetic resonance imaging (MRI) scans from the BraTS 2020 challenge, as well as from various remote independent institutions included in the collaborative network of a real-world federation\footnote{\url{https://www.fets.ai/}}. The goals of the FeTS challenge are directly represented by the two included tasks: 1) the identification of the optimal weight aggregation approach towards the training of a consensus model that has gained knowledge via federated learning from multiple geographically distinct institutions, while their data are always retained within each institution, and 2) the federated evaluation of the generalizability of brain tumor segmentation models ``in the wild'', i.e. on data from institutional distributions that were not part of the training datasets.
\end{abstract}


% =========================
% main part of the document
% =========================
\section{Introduction}
Glioblastomas (GBM) are arguably the most aggressive and heterogeneous adult brain tumor. In spite of the proliferation of multimodal treatment composed of maximal safe surgical resection, radiation and chemotherapy, the median survival has only slightly improved to approximately 15 months, with less than 10\% of patients surviving for over 5 years \cite{cbtrus_2019}. This tremendously poor prognosis is largely on account of the pathological heterogeneity inherently present in GBM tumors, leading to treatment resistance, and thus woeful patient outcomes. Radiologic imaging (i.e., magnetic resonance imaging (MRI)) is the modality of choice for routine clinical diagnosis and response assessment in GBM, and delineation of the tumor sub-regions is the first step towards any computational analysis that can enable personalized diagnostics \cite{pati2020reproducibility}.
    
To enable robust tumor delineation, the Brain Tumor Segmentation (BraTS) challenge \cite{menze2014multimodal,bakas2018identifying,bakas2017advancing} has been at the forefront in providing widely-used good quality benchmarking data and in making algorithms available for use by the scientific community \cite{bakas2015glistrboost,zeng2016segmentation,kamnitsas2017efficient,isensee2018nnu,mckinley2018ensembles}. However, to ensure that the greatest breadth of data available for generalizable model construction, additional patient datasets from geographically distinct locations need to be made available. Unfortunately, due to regulatory and privacy hurdles \cite{hipaa,gdpr}, collection of such data in a single site is virtually impossible.

The Federated Tumor Segmentation (FeTS) challenge \cite{spyridon_bakas_2021_4573128} is the first challenge to ever be proposed for Federated Learning (FL) and intends to address the hurdles posed by data privacy and regulatory concerns by leveraging FL \cite{mcmahan2017communication,fl_1,fl_2,fl_3,sheller2020federated}. Specifically, the FeTS 2021 challenge uses clinically acquired, multi-parametric MRI (mpMRI) scans from the multi-institutional BraTS 2020 dataset \cite{menze2014multimodal,bakas2018identifying,bakas2017advancing,bakas2017segmentation_1,bakas2017segmentation_2}, as well as data from various remote independent institutions included in the collaborative network of a real-world federation\footnote{\url{https://www.fets.ai/}}. The FeTS challenge focuses on the construction and evaluation of a consensus model for the segmentation of intrinsically heterogeneous (in appearance, shape, and histology) brain tumors, namely gliomas \cite{sheller2020federated}. The ultimate goal of FeTS is two-fold:
    \begin{enumerate}
    \item the generation of a consensus segmentation model that has gained knowledge from data of multiple institutions without pooling their data together (i.e., by retaining the data within each institution), and
    \item the evaluation of segmentation models in a federated configuration (``in the wild''), to assess their robustness to dataset shifts in multi-institutional data.
\end{enumerate}
These goals are reflected in the two challenge tasks, introduced in \cref{sec:task1,sec:task2}. The present manuscript further describes the mission and design of the FeTS 2021 Challenge. We intend to extend it with the results of the challenge in a future version. A structured challenge design document, describing the aspects of the challenge according to the BIAS protocol \cite{maier2020bias}, to ensure consistent interpretation of the challenge results, is available at \cite{spyridon_bakas_2021_4573128}.

\section{Materials \& Methods}
    \subsection{Challenge Datasets}
\label{sec:data}

    \subsubsection{Data Sources}
     This challenge leverages data from two sources: the BraTS 2020 challenge data \cite{menze2014multimodal,bakas2018identifying,bakas2017advancing}, as well as data from collaborators in the FeTS federation \footnote{\url{https://www.fets.ai/}}. The following sections apply to both of them if not noted explicitly. Both sources contain clinically acquired mpMRI along with their ground truth annotations for the tumor sub-regions, augmented with meta-data that identify the origin of each scan and its corresponding acquisition protocols. Each case describes mpMRI scans for a single patient at the pre-operative baseline timepoint. The exact mpMRI sequences included for each case are i) native (T1) and ii) contrast-enhanced T1-weighted (T1-Gd), iii) T2-weighted (T2), and iv) T2 Fluid Attenuated Inversion Recovery (T2-FLAIR). 
     
     \subsubsection{Data Pre-processing}
         The exact pre-processing pipeline applied to all the data considered in the FeTS 2021 challenge is identical with the one evaluated and followed by the BraTS challenge. Specifically, all input scans (i.e., T1, T1-Gd, T2, T2-FLAIR) are rigidly registered to the same anatomical atlas (i.e., SRI-24 \cite{sri}) using the Greedy diffeomorphic registration algorithm \cite{yushkevich2016fast}, ensuring a common spatial resolution of $(1 mm^3)$. After completion of the registration process, brain extraction is done to remove any apparent non-brain tissue, using a deep learning approach specifically designed for brain MRI scans with apparent diffuse glioma. This algorithm utilizes a novel training mechanism that introduces the brain's shape prior as knowledge to the segmentation algorithm \cite{thakur2020brain}. All pre-processing routines have been made publicly available through the Cancer Imaging Phenomics Toolkit (CaPTk\footnote{\url{https://www.cbica.upenn.edu/captk}}) \cite{captk_1,captk_2,captk_3} and FeTS \footnote{\url{https://www.fets.ai/}}.
         
    \subsubsection{Annotation Protocol}
         The skull-stripped scans are then used for annotating the brain tumor sub-regions. The annotation process follows a pre-defined clinically approved annotation protocol, which was provided to all clinical annotators, describing in detail the radiologic appearance of each tumor sub-region according to the specific provided MRI sequences. The annotators were given the flexibility to use their tool of preference for making the annotations, and also follow either a complete manual annotation approach, or a hybrid approach where an automated approach is used to produce some initial annotations followed by their manual refinements. The protocol followed by annotators is given below: 

        \begin{enumerate}
            \item the enhancing tumor (ET) delineates the hyperintense signal of the T1-Gd, after excluding the vessels.
            \item the necrotic tumor core (NCR) outlines regions appearing dark in both T1 and T1-Gd images (denoting necrosis/cysts), and dark regions in T1-Gd and bright in T1.
            \item the tumor core (TC) includes the ET and NCR, and represents what is typically resected during a surgical operation.
            \item the farthest tumor extent (what is also called the whole tumor) including the peritumoral edematous and infiltrated tissue (ED) and delineates the regions characterized by the hyperintense abnormal signal envelope on T2 \& T2-FLAIR sequences.
        \end{enumerate}
        The provided segmentation labels have values of $1$ for NCR, $2$ for ED, $4$ for ET, and $0$ for everything else.
        
        For the BraTS-data, each case was assigned to a pair of annotator-approver. Annotators spanned across various experience levels and clinical/academic ranks, while the approvers were the 2 experienced board-certified neuroradiologists (with $\geq$ 13 years of experience with glioma). The annotators were given the flexibility to use their tool of preference for making the annotations, and also follow either a complete manual annotation approach, or a hybrid approach where an automated approach is used to produce some initial annotations followed by their manual refinements. Once the annotators were satisfied with the produced annotations, they were passing these to the corresponding approver. The approver is then responsible for signing off these annotations. Specifically, the approver would review the tumor annotations, in tandem with the corresponding mpMRI scans, and if the annotations were not of satisfactory quality they would be sent back to the annotators for further refinements. This iterative approach was followed for all cases, until their respective annotations reached satisfactory quality (according to the approver) for being publicly available and noted as final ground truth segmentation labels for these scans.
        
        Collaborators from the FeTS federation were free to choose their preferred annotation workflow, but we recommended to use the following semi-automatic approach based on the predictions of an ensemble of state-of-the-art BraTS models provided to them. Specifically, a tool was passed to the FeTS federation collaborators containing pre-trained models of the DeepMedic \cite{kamnitsas2017efficient}, nnU-Net \cite{isensee2018nnu}, and DeepScan \cite{mckinley2018ensembles} approaches trained on the BraTS data, with their label fusion performed using the Simultaneous Truth and Performance Level Estimation (STAPLE) algorithm \cite{warfield2002validation,labelfusion}. Refinements of the fused labels were then performed by the clinical experts at each of the collaborating sites according to the BraTS annotation protocol \cite{bakas2018identifying}. Sanity checks were performed to ensure the integrity of annotations.
        
    \subsubsection{Training, Validation, and Test case characteristics}

    Training and Validation sets are cases from the BraTS 2020 dataset. Out of these, only the subset of radiographically appearing glioblastoma are included in the FeTS challenge, and cases without an apparent enhancement were excluded. %A total of $n_{train}=XXX$  training and $n_{val} = YYY$ cases are used in both tasks of the FeTS challenge.
    Training cases encompass the mpMRI volumes, the corresponding tumor sub-region annotations, as well as meta-data that identify the site where the scans were acquired. In contrast, validation cases only contain the mpMRI volumes, without any accompanying ground truth annotations.
        
    The testing datasets are from BraTS 2020 and from the FeTS federation collaborators. Notably, these are not shared with the challenge participants or the public. %For Task-2, data from the FeTS federation will be used instead of the BraTS-based test set.
    The FeTS collaborators collect and process data from their clinical sites following the pre-processing and annotation protocols presented above. From all these data, approximately 80\% is kept for federated training (outside the context of this challenge) and the rest is set aside for testing and that will be used to evaluate the performance of submitted algorithms. Note that the corresponding mpMRI scans are not shared with the organizers and have to remain decentralized during challenge evaluation.
    
\subsection{Performance Evaluation}
\label{sec:eval_metrics}

    This section describes the procedure for evaluating the performance of participating algorithms. Participants are called to produce segmentation labels of the different glioma sub-regions:
    \begin{enumerate}
        \item the ``enhancing tumor'' (ET), equivalent to label 4,
        \item the ``tumor core'' (TC), comprising labels 2 and 4,
        \item the ``whole tumor'' (WT), comprising labels 1, 2 and 4
    \end{enumerate}
    For each region, the predicted segmentation is compared with the ground truth segmentation using the following metrics:

     \subsubsection*{Dice similarity coefficient (DSC)}
        The DSC is a metric commonly used to evaluate the performance of segmentation tasks. It measures the extent of spatial overlap, while taking into account the intersection between the predicted masks ($PM$) and the provided ground truth ($GT$), hence handles over- and under-segmentation, and is defined as
     
        \begin{equation}
        \label{eq:dice}
            DSC = \frac{2|GT \cap PM|}{|GT|+|PM|} \, .
        \end{equation}

     \subsubsection*{Hausdorff distance (HD)}
        This metric quantifies the distance between the boundaries of the ground truth labels against the predicted label. It originated from set theory and measures the maximum distance of a point set to the nearest point in another set \cite{hausdorff}. This makes the HD sensitive to local differences, as opposed to the DSC, which represents a global measure of overlap. For the specific problem of brain tumor segmentation, local differences may be very important for properly assessing the quality of the segmentation.
     In this challenge, the 95\textsuperscript{th} percentile of the \textit{Hausdorff} distance between the contours of the two segmentation masks is calculated, which is a variant of HD that is more robust to outlier pixels:
        \begin{equation}
        \label{eq:haus}
            H_{95}(PM, GT)  = \max \left\{
            \underset{p \in PM}{P_{95\%}}\,  d(p, GT), \underset{g \in GT}{P_{95\%}}\, d(g, PM)   \right\}\, , 
        \end{equation}
    where $d(x, Y) = \min_{y\in Y} ||x - y||$ is the distance of $x$ to set $Y$.

    \subsubsection*{Communication cost during model training}
    The ``cost'' represents the budget time, which is the product of number of bytes sent/received multiplied by number of federated rounds. This metric will be used to assess the performance of the weight aggregation methods in Task 1 only.
                 
     \iffalse
     \item Sensitivity (for additional information) 
     \item Specificity (for additional information)  
     \fi

    \subsection{Code Availability}
    
        Towards encouraging reproducibility by the community, all tools, pipelines, and methods have been released through the Cancer Imaging Phenomics Toolkit (CaPTk) \cite{captk_1,captk_2,captk_3} and the Federated Tumor Segmentation tool\footnote{\url{https://github.com/FETS-AI/Front-End/}}. Challenge-specific instructions and code to produce the rankings are publicly available\footnote{\url{https://github.com/FETS-AI/Challenge}}.

    \subsection{Task 1: Federated Training (FL Weight Aggregation Methods)}
\label{sec:task1}

\subsubsection{Mission}
    The first task of the challenge involves creating a robust consensus model for segmentation of brain tumor sub-regions that has gained knowledge from data acquired at multiple sites (see Section \ref{sec:data}), without pooling data together. The specific focus of this task is to identify the best way to aggregate the knowledge coming from segmentation models trained on individual institutions, instead of identifying the best segmentation method. More precisely, the focus is on the methodological portions specific to federated learning (e.g., aggregation, client selection, training-per-round, compression, communication efficiency), and not on the development of segmentation algorithms (which is the focus of the BraTS challenge). To facilitate this, an existing infrastructure for federated tumor segmentation using federated averaging will be provided to all participants indicating the exact places that the participants are allowed and expected to make changes. The primary objective of this task is to develop methods for effective aggregation of local segmentation models, given the partitioning of the data into their real-world distribution. As an optional sub-task, participants will be asked to account for network communication outages, i.e., dealing with stragglers.

% The first task of the challenge involves creating a robust consensus model for segmentation of brain tumor sub-regions that has gained knowledge from data acquired at multiple sites (see Section \ref{sec:data}), without pooling data together.

\subsubsection{Model Architecture}
\label{sec:arch}
\begin{figure}
        \centering
        \includegraphics[width=1\textwidth]{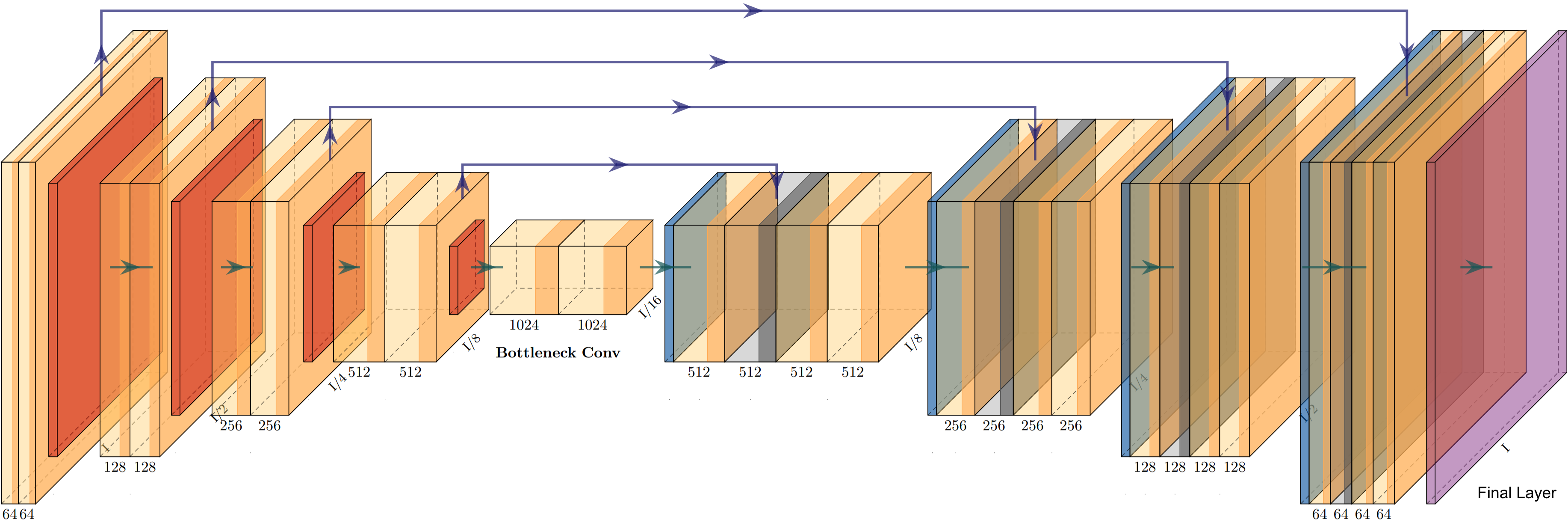}
        \caption{The U-Net architecture. Figure was plotted using PlotNeuralNet \cite{iqbal2018harisiqbal88}.}
        \label{fig:unet}
    \end{figure}

    To focus on the development of aggregation methods, we needed to fix the architecture of the segmentation model. Based on the current literature, the architecture of the segmentation model evaluated in this task will be a U-Net \cite{unet} (Figure. \ref{fig:unet}) with residual connections, which has shown to provide a robust performance for medical imaging datasets \cite{isensee2018nnu,thakur2020brain,drozdzal2016importance,he2016deep,cciccek20163d,pati2021gandlf}. The U-Net consists of an encoder, comprising convolutional layers and downsampling layers, and a decoder offering upsampling layers (applying transpose convolution layers) and includes skip connections in every convolution block. The encoder-decoder structure contributed in automatically capturing information at multiple scales/resolutions. The U-Net further includes skip connections, which consist of concatenated feature maps paired across the encoder and the decoder layer, to improve context and feature re-usability. The residual connections utilize additional information from previous layers (across the encoder and decoder) that enables a boost in the performance. 
    
\subsubsection{Model Aggregation}
\label{sec:aggregation}
\begin{figure}
        \centering
        \includegraphics[width=0.75\textwidth]{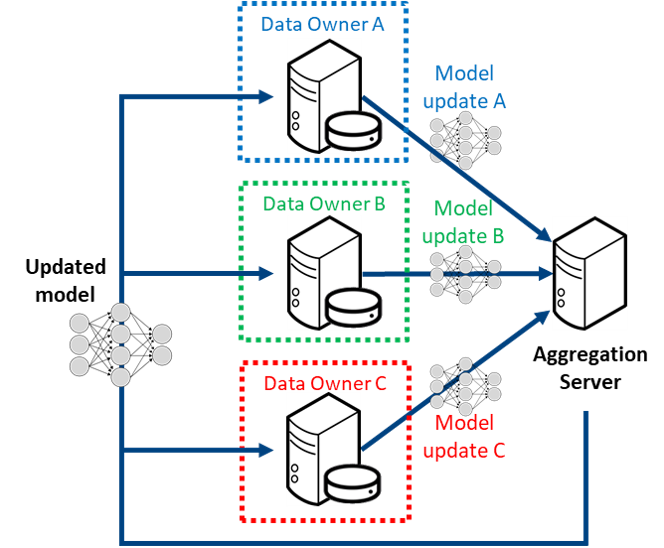}
        \caption{The model aggregation mechanism during FL-based training \cite{sheller2018multi,sheller2020federated}.}
        \label{fig:agg}
    \end{figure}

    Following extensive prior literature \cite{sheller2018multi,sheller2020federated,isensee2018nnu,pati2021gandlf}, the final model for each local individual institutional training is taken as the one that achieves the best local validation score over the course of training. During a single round of FL-based training \cite{sheller2018multi,sheller2020federated,fl_3} (illustrated by Figure \ref{fig:agg}), each partnering institution (hereby named a \textit{collaborator}) locally validates any model it receives from the central aggregation server (hereby named the \textit{aggregator}), i.e., at the start of each federated round. Each \textit{collaborator} then trains the model received from the \textit{aggregator} on their local data to update the model gradients. The local validation results along with the model updates of each site are then sent to the \textit{aggregator}, which combines all the model updates to produce a new consensus model. The new consensus model is then passed back to each \textit{collaborator}, and starting a new round of federation.
    
    \subsection{Task 2: Federated Evaluation (Inference ``In The Wild'')}
\label{sec:task2}

\subsubsection{Mission}
    The discrepancy between AI systems' performance in research environments and real-life applications is one of the key challenges in our field \cite{nagendran2020artificial,zech2018variable,albadawy2018deep,badgeley2019deep,beede2020human}. This ``AI chasm'' can be attributed in part to the limited diversity of training datasets, which do not necessarily reflect the variety of real-world datasets ``in the wild''. As a consequence, most deep learning models exhibit limited generalizability when applied to datasets acquired from different imaging devices and populations. Federated setups are not only beneficial for learning models: they also allow to extend the size and diversity of typical test datasets substantially, as clinicians may contribute data to a challenge without having to publicly release them, thus constituting an important step towards the evaluation of model robustness in the wild.
    
    In this task, the goal is to find algorithms that robustly produce accurate brain tumor segmentations across different medical institutions, MRI scanners, image acquisition parameters and populations. To this end, we use a real-world federated evaluation environment (based on the FeTS federation\footnote{\url{https://www.fets.ai/}}), which enables scaling up algorithm validation in the context of competitions. Based on the wide variety of data distributions covered by the federation, we hope to gain insights into which i) algorithmic components are crucial for robustness and ii) potentially which dataset characteristics might hinder generalization.

\subsubsection{Organization}
    In the training phase, the participants will be provided the training set including information on the data origin, described in \cref{sec:data}. They can explore the effects of data partitioning and distribution shifts between contributing sites, with the aim of finding tumor segmentation algorithms that are able to generalize to data acquired at institutions that did not contribute to the training dataset. Note that \textit{training on pooled data will be allowed} in this task, so that the participants can develop methods that optimally exploit the meta-information of data origin.
    
    In the validation phase, participants can evaluate their model on the validation set described in \cref{sec:data} to estimate in-distribution generalization. However, since task-2 is about out-of-distribution generalization, caution should be exercised when selecting the best performing model purely from in-distribution data. As indicated by \cite{gulrajani2021in}, model selection is an important design choice and provides further research opportunities to the participants.
    
    After training, all participating algorithms will be evaluated in a distributed way on datasets from various institutions of the FeTS federation (see \cref{sec:data}), such that the test data are always retained within their owners’ servers. Singularity containers \cite{kurtzer2017singularity}\footnote{\url{https://sylabs.io/singularity/}} were chosen as the submission format, as they provide the participants with high flexibility and allow running algorithms securely on different systems.

\subsubsection{Methodological Context: Domain Generalization}\label{sec:domain_general}
     The central characteristic of this task is the distribution shift between training/validation and test set, which naturally arises from the differences between real-world medical institutions. Hence, as each institution can be associated with a different data distribution (``domain''), task-2 can be viewed as a \textit{domain generalization} (DG) problem: Given data from multiple training domains, the goal is to learn a model that generalizes well to other domains not seen during training \cite{blanchard2011generalizing}. While DG is far from solved \cite{gulrajani2021in}, promising results have been reported in the medical field using data augmentation \cite{zhang2020generalizing,full2020studying}, meta-learning \cite{dou2019domain,liu2020shape} and test-time adaptation \cite{karani2021test}. Pioneering work has even incorporated restrictions of an FL setup into a DG method \cite{liu2021feddg}, which we consider an interesting future direction.
\section{Discussion}

Here we present the first ever computational competition (i.e., challenge) in FL. We particularly focus on a healthcare application, where privacy is a major concern. Stringent legal regulations such as the Health Insurance Portability and Accountability Act (HIPAA) of the United States of America \cite{hipaa} and General Data Protection Regulation (GDPR) of the European Union \cite{gdpr} ensure protection for patients, but they introduce major hurdles towards constructing common datasets that can further scientific research.

Current challenges in the field of medicine include limited multi-institutional data, which hinders the ability to train robust and generalizable models. FL provides a mechanism to potentially instigate a paradigm shift in the way multi-site collaborations are looked at by enabling model training without the requirement to share data. This could serve regulatory and privacy concerns while providing ample avenue for scientific research.

While federated evaluation schemes circumvent many of the common obstacles with sharing medical data (such as data-privacy issues), they also come with their own set of challenges and particularities. In particular, setting up and coordinating a federation with heterogeneous regulations is still a demanding task and requires communication and compromises. Another aspect is data harmonization and quality control. Standardized pre-processing, as well as tests and sanity checks are required to ensure that the data can be used to develop and benchmark algorithms. Importantly, despite identical annotation protocols, inconsistencies in the reference segmentations are likely to occur in a federated setting with several different annotators. A basic countermeasure we employed is a semi-automatic annotation workflow, where manual refinement of segmentations produced by state-of-the-art algorithms encourages the annotations towards a common style. By addressing these obstacles, we hope that this challenge can provide a blueprint for similar endeavors in the future, where successful challenges could enter a “phase 2” with a federated setup and thus move one step closer towards the real-life use case.

Past FL studies \cite{fl_1,fl_2,fl_3,sheller2020federated,sheller2018multi,sarma_federated_2021,roth_federated_2020,remedios_distributed_2020,wang_automated_2020} have shown significant promise towards the training of robust DL models, but they have either conducted a simulated federation, or had a handful  of real-world collaborators. Preliminary results from applications using large-scale, real-world, geographically distinct collaborations \footnote{\url{https://www.fets.ai/}} look promising, and provide a window into a paradigm where data silos can be maintained without blocking computational research. Ideally, the methods from both challenge tasks, federated learning and domain generalization, will complement each other to accelerate the translation of advanced computational algorithms into clinical practice. 

We hope that with this challenge, FeTS can engage the computational community towards more FL-based research, while providing the building blocks and a common platform to initiate future studies based on FL.

\iffalse % kept commented because I guess we want to populate this after we have some results?
\section{Conclusion}
    
    Text related to the discussion goes here.
\fi
% =========================
\iffalse
\section*{Author Contributions}
    Study conception and design: 
    Software development used in the study: 
    Wrote the paper: 
    Data analysis and interpretation: 
    Reviewed / edited the paper: 
\fi

\section*{Acknowledgments}
    Research reported in this publication was partly supported by the National Institutes of Health (NIH) under award numbers NIH/NCI:U01CA242871, NIH/NINDS:R01NS042645, and NIH/NCI:U24CA189523. This work is also partly funded by the Helmholtz Association (HA) within the project ``Trustworthy Federated Data Analytics'' (TFDA) (funding number ZT-I-OO1 4). The content of this publication is solely the responsibility of the authors and does not represent the official views of the NIH or the HA. 

\bibliographystyle{ieeetr}
\bibliography{bibliography.bib}

\end{document}